\documentclass[english,prl, twocolumn]{revtex4}
\usepackage[T1]{fontenc}
\usepackage[latin9]{inputenc}
\usepackage{color}
\usepackage{babel}
\usepackage{amssymb}
\usepackage{graphicx}
\usepackage[unicode=true,pdfusetitle,
 bookmarks=true,bookmarksnumbered=false,bookmarksopen=false,
 breaklinks=false,pdfborder={0 0 1},backref=false,colorlinks=false]
 {hyperref}

\makeatletter

\newcommand*\LyXThinSpace{\,\hspace{0pt}}

\@ifundefined{textcolor}{}
{%
 \definecolor{BLACK}{gray}{0}
 \definecolor{WHITE}{gray}{1}
 \definecolor{RED}{rgb}{1,0,0}
 \definecolor{GREEN}{rgb}{0,1,0}
 \definecolor{BLUE}{rgb}{0,0,1}
 \definecolor{CYAN}{cmyk}{1,0,0,0}
 \definecolor{MAGENTA}{cmyk}{0,1,0,0}
 \definecolor{YELLOW}{cmyk}{0,0,1,0}
}

\makeatother

\begin{document}

\title{Renormalized stress-energy tensor of an evaporating spinning black
hole}

\author{Adam Levi\textsuperscript{1}, Ehud Eilon\textsuperscript{1}, Amos
Ori\textsuperscript{1} and Maarten van de Meent\textsuperscript{2}}

\address{\textsuperscript{1}Department of physics, Technion-Israel Institute
of Technology, Haifa 32000, Israel. \\
\textsuperscript{2}Mathematical Sciences, University of Southampton,
Southampton SO17 1BJ, United Kingdom.}
\begin{abstract}
We employ a recently developed mode-sum regularization method to compute
the renormalized stress-energy tensor of a quantum field in the Kerr
background metric (describing a stationary spinning black hole). More
specifically, we consider a minimally-coupled massless scalar field
in the Unruh vacuum state, the quantum state corresponding to an evaporating
black hole. The computation is done here for the case $a=0.7M$, using
two different variants of the method: $t$-splitting and $\varphi$-splitting,
yielding good agreement between the two (in the domain where both
are applicable). We briefly discuss possible implications of the results
for computing semiclassical corrections to certain quantities, and
also for simulating dynamical  evaporation of a spinning black hole.
\end{abstract}
\maketitle
Like many great discoveries, Hawking's discovery of black-hole (BH)
evaporation \cite{Hawking - Particle creation by black holes} opened
a number of new profound questions.  Two such outstanding questions
are the information loss puzzle, and \textemdash{} more generally
\textemdash{} what is the end state of BH evaporation. A closely related
problem is the detailed study of the semiclassical evaporation process.
Hawking's original analysis uses quantum field theory in curved spacetime
to determine the flux that a BH emits to infinity. However, a more
detailed investigation of BH evaporation requires not just the outflux
at infinity but also the full renormalized stress-energy tensor (RSET)
$\left\langle T_{\alpha\beta}\right\rangle _{ren}$; namely, the contribution
of the quantum field fluctuations to the local stress-energy tensor.
It can then be inserted in the semiclassical Einstein equation
\[
G_{\alpha\beta}=8\pi\left\langle T_{\alpha\beta}\right\rangle _{ren}
\]
to investigate the back-reaction on the metric. Here $G_{\alpha\beta}$
is the Einstein tensor, and throughout this paper we use general-relativistic
units $G=c=1$, along with $\left(-+++\right)$ signature.

The calculation of the RSET is a long-standing challenge, even for
a prescribed background metric. The naive quantum-field computation
yields a divergent mode sum. To renormalize it one can use the point-splitting
procedure, originally developed by DeWitt \cite{Dewitt - Dynamical theory of groups and fields}
for $\left\langle \phi^{2}\right\rangle _{ren}$ and later adjusted
to the RSET by Christensen \cite{Christiansen}. The point-splitting
scheme proves to be very useful when the field modes are known analytically.
However in our case of interest \textemdash{} BH backgrounds \textemdash{}
the field's modes are known only numerically, making the naive implementation
of the scheme impractical. To overcome this difficulty, Candelas,
Howard, Anderson and others \cite{Candelas =000026 Howard - 1984 - phi2 Schwrazschild,Howard - 1984 - Tab Schwarzschild,Anderson - 1990 - phi2 static spherically symmetric,Anderson - 1995 - Tab static spherically symmetric}
developed a method to implement point-splitting numerically. This
method requires a fourth order WKB expansion. Since performing high-order
WKB expansion is extremely difficult in Lorentzian spacetime, they
used Wick rotation and carried the actual calculation in the Euclidean
sector. This clever trick is very restrictive however, as the Euclidean
sector does not generically exist. The most general case where this
method was implemented is a static spherically-symmetric background
\cite{Anderson - 1995 - Tab static spherically symmetric}. \footnote{An interesting new method (still for spherical static backgrounds)
was recently proposed by Taylor and Breen \cite{Peter taylor}.}

Note also that on going to the Euclidean sector one cannot compute
the RSET directly in Unruh state, as the latter is not defined there.
Instead, one has to compute another state (e.g. Boulware) in the Euclidean
sector, and then use the technique introduced by Elster \cite{Elster}
to compute the difference between two states (a non-divergent quantity)
in the Lorentzian sector. This method was used to compute the RSET
in Schwarzschild in Unruh state for conformally-coupled scalar field
\cite{Elster} and also for electromagnetic field \cite{Ottewill - Schwarzschild Electromagnetic}.

The traditional method \cite{Candelas =000026 Howard - 1984 - phi2 Schwrazschild,Howard - 1984 - Tab Schwarzschild,Anderson - 1990 - phi2 static spherically symmetric,Anderson - 1995 - Tab static spherically symmetric}
is inapplicable to the Kerr geometry (describing a spinning BH) which
is neither spherically-symmetric nor static \textemdash{} and does
not admit a Euclidean sector. Spinning BH solutions are considerably
more realistic than static ones, because astrophysical BHs do generally
rotate. A fascinating demonstration of this fact was recently provided
by the two gravitational-wave signals discovered by LIGO: \cite{LIGO summary paper}
In both merger events GW150914 and GW151226 the final BHs were significantly
spinning (with $a/M\sim0.7$). 

The above discussion highlights the importance of generalizing the
methods of RSET computation from spherical static BHs to the Kerr
case. Over the years Ottewill, Casals, Winstanley, Duffy and others
made remarkable progress by posing various quantum states on the Kerr
metric \cite{Ottewill - Kerr - General results,Ottewill - Kerr - electromagnetic differences},
and also by computing RSET differences between pairs of quantum states
\cite{Ottewill - Kerr - electromagnetic differences,Ottewill - Kerr with mirror}
(which are regular). Another approach was to compute the RSET for
rotating BHs in 2+1D \cite{Casals Fabbri Mart=0000EDnez Zanelli - 2016 - 2+1 BTZ,Hugo R. C. Ferreira}
as a toy model for 3+1D.

Recently, a novel approach for implementing point-splitting and computing
$\left\langle \phi^{2}\right\rangle $ and $\left\langle T_{\alpha\beta}\right\rangle _{ren}$
in BH spacetimes was introduced by Levi and Ori \cite{Levi =000026 Ori - 2015 - t splitting regularization,Levi =000026 Ori - 2016 - theta splitting regularization,Levi =000026 Ori - 2016 - RSET},
to which we shall refer here as the ``Pragmatic Mode-sum Regularization''
(PMR) method. This method does not resort to the Euclidean sector
or to WKB expansion. Basically it only requires the background to
admit a single (continuous) symmetry. PMR comes in several variants,
depending on the symmetry being employed. So far two variants were
introduced in detail, the $t$-splitting \cite{Levi =000026 Ori - 2015 - t splitting regularization}
and angular-splitting \cite{Levi =000026 Ori - 2016 - theta splitting regularization}
variants, applicable to stationary or spherically-symmetric backgrounds
respectively. For the sake of simplicity the presentation in \cite{Levi =000026 Ori - 2015 - t splitting regularization,Levi =000026 Ori - 2016 - theta splitting regularization}
was restricted to the regularization of $\left\langle \phi^{2}\right\rangle $,
which is technically simpler. A third variant, $\varphi$-splitting
(also named ``azimuthal splitting''), aimed for axially-symmetric
backgrounds, was also briefly introduced, in a very recent paper \cite{Levi =000026 Ori - 2016 - RSET}
which presented the RSET computation in Schwarzschild using PMR. All
three variants were used in that paper, showing very good agreement
between the three splittings.

Because PMR requires only one symmetry, it can actually be used to
compute the RSET in Kerr in two different ways, once using $t$-splitting
and once using $\varphi$-splitting. The former primarily relies on
the field decomposition in temporal modes $e^{-i\omega t}$, and the
latter on (discrete) decomposition in azimuthal modes $e^{im\varphi}$.
Having two independent regularization variants is advantageous as
it allows one to test the method's consistency as well as numerical
accuracy. Moreover, each splitting variant breaks down in a certain
locus, where the norm of the associated Killing field vanishes. This
happens to $t$-splitting at the ergosphere boundary, and to $\varphi$-splitting
at the polar axis. In reality, the splitting variant becomes problematic
also in some neighborhood of that singular locus. Using the two variants
allows one to compute the RSET almost everywhere outside the BH. 

In this Letter we present the results obtained (from both $t$-splitting
and $\varphi$-splitting) for $\left\langle \phi^{2}\right\rangle _{ren}$
and $\left\langle T_{\alpha\beta}\right\rangle _{ren}$ in Kerr background,
for a minimally-coupled massless scalar field in Unruh state \cite{Unruh - Unruh state}
\textemdash{} the quantum state representing an evaporating BH.

\paragraph{Kerr metric and modes computation.\textendash{}\emph{ }}

The Kerr metric in Boyer-Lindquist coordinates is 
\[
\begin{array}{c}
ds^{2}=-\frac{\Delta}{\Sigma}\left(dt-a\sin^{2}\theta\,d\varphi\right)^{2}+\frac{\Sigma}{\Delta}dr^{2}+\Sigma d\theta^{2}\\
+\frac{\sin^{2}\theta}{\Sigma}\left[\left(r^{2}+a^{2}\right)d\varphi-a\,dt\right]^{2}
\end{array}
\]
where $\Delta\equiv r^{2}-2Mr+a^{2}$ and $\Sigma\equiv r^{2}+a^{2}\cos^{2}\theta$,
$M$ is the BH mass and $a$ its angular momentum per unit mass. We
have chosen to work here on the case $a=0.7M$, which is strongly
motivated by the BH merger outcomes in the two recent LIGO detections
\cite{LIGO summary paper}. The field modes were constructed according
to the boundary conditions formulated by Ottewill and Winstanley \cite{Ottewill - Kerr - General results}.
The computation was done by solving the spin-0 Teukolsky equation
using a numerical implementation \cite{Maarten-MST} of the Mano-Suzuki-Takasugi
(MST) formalism \cite{MST 1,MST 2}. Modes were computed for $-60\leq m\leq60$
and for $\omega$ from zero to $\omega_{max}=8M^{-1}$ with uniform
spacing of $0.01M^{-1}$. For each $\omega$ and $m$ the sum over
$l$ was preformed until sufficient convergence was achieved. In total,
just over 4 million $lm\omega$-modes were used.

\paragraph*{Results for $\left\langle \phi^{2}\right\rangle _{ren}$ in Kerr.\textendash{} }

In calculating $\left\langle \phi^{2}\right\rangle _{ren}$ using
$t$-splitting \cite{Levi =000026 Ori - 2015 - t splitting regularization},
one has to integrate a certain function $F_{reg}(\omega)$ over $\omega$.
This function contains oscillations, originating from ``connecting
null geodesics'' (CNGs), \cite{Levi =000026 Ori - 2015 - t splitting regularization}
whose wavelengths (in $\omega$) are dictated by the length (in $t$)
of these CNGs. The self-cancellation method introduced in Ref. \cite{Levi =000026 Ori - 2015 - t splitting regularization}
to eliminate the oscillations requires knowledge of these wavelengths.
In the Schwarzschild case we determined them by numerically finding
the CNGs. In Kerr, however, it is a bit more difficult to compute
the CNGs. We therefore used two alternative techniques. The first
was finding the wavelengths from a Fourier transform of $F_{reg}(\omega)$,
and self-canceling the oscillations according to the recipe of \cite{Levi =000026 Ori - 2015 - t splitting regularization}.
The second was simply to apply a low-pass filter to $F_{reg}(\omega)$,
to eliminate the oscillations. The two techniques produced very similar
results.

We also computed $\left\langle \phi^{2}\right\rangle _{ren}$ using
$\varphi$-splitting. This variant, which was used recently \cite{Levi =000026 Ori - 2016 - RSET}
for the computation of $\left\langle T_{\alpha\beta}\right\rangle _{ren}$
in Schwarzschild, will be presented in detail elsewhere \cite{Levi =000026 Ori - In preparation}.
We should mention, briefly, that in $\varphi$-splitting regularization
we first sum the $lm\omega$ mode contributions over $l$ (a convergent
sum), to obtain the functions $F(\omega;m)$. Next we regularize the
integrals of $F(\omega;m)$ over $\omega$ (for each $m$), and finally
we regularize the sum over $m$. Here, too, one finds that $F(\omega;m)$
exhibits oscillations in $\omega$. In some analogy with the angular-splitting
case \cite{Levi =000026 Ori - 2016 - theta splitting regularization},
these oscillations originate from CNGs in a fictitious 2+1 dimensional
spacetime (obtained from the Kerr metric by eliminating the $\varphi$
coordinate). We were able to numerically compute these reduced-dimension
CNGs and to obtain the oscillations' wavelengths, which we then used
to self-cancel the oscillations. After integrating the (smoothened
and regularized) functions $F(\omega;m)$ over $\omega$, the sum
over $m$ is regularized using a discrete Fourier decomposition of
the counter-terms, in analogy to the Fourier decomposition of the
latter in $t$-splitting \cite{Levi =000026 Ori - 2015 - t splitting regularization}. 

Figure \ref{fig: Figure 1} displays the results for $\left\langle \phi^{2}\right\rangle _{ren}$
versus $\theta$ (for various $r$ values) in Unruh state, for $a/M=0.7$,
obtained using both variants: $t$-splitting in solid curves, and
$\varphi$-splitting in ``+'' symbols. Here, and also in all other
figures, numerical results are given in units $M=1$ (in addition
to $G=c=1$). Note that for $\varphi$-splitting we only give results
for $\theta\geq30^{\circ}$ because its accuracy rapidly deteriorates
on getting closer to the pole. \footnote{In fact the relevant parameter is $r\sin\theta$.}
The agreement between the two variants is better than one part in
$10^{3}$ throughout the domain presented. This agreement steadily
improves with increasing $\theta$: At $\theta\geq50^{\circ}$ it
is better than two parts in $10^{5}$, and for $\theta=90^{\circ}$
it is about one part in $10^{6}$. We can independently estimate the
accuracy of our $t$-splitting results, it is usually better than
one part in $10^{6}$ (throughout $4M\le r\le10M$). Therefore, throughout
the domain shown in Fig. \ref{fig: Figure 1} we may associate the
disagreement between the two variants with the inaccuracy in $\varphi$-splitting.
\footnote{This situation is reversed in the neighborhood of $r\approx2M,\theta>1$
(not included in Fig. \ref{fig: Figure 1}).} 

\begin{center}
\begin{figure}[h]
\centering{}\includegraphics[bb=30bp 180bp 550bp 650bp,clip,scale=0.45]{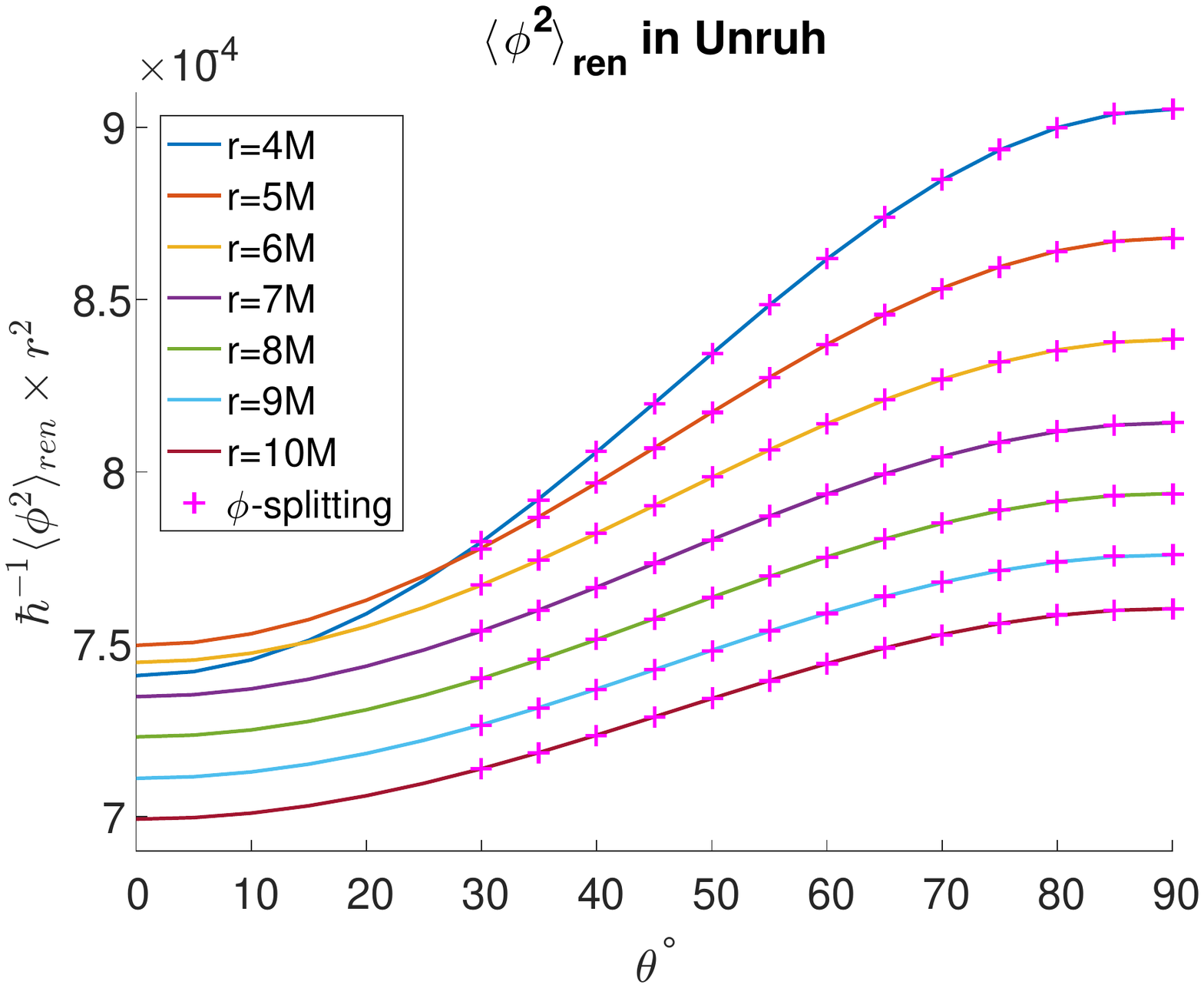}\caption{Results for $\left\langle \phi^{2}\right\rangle _{ren}\times r^{2}$
in Unruh state in Kerr, from both $t$-splitting (solid curves) and
$\varphi$-splitting (``+'' symbols). \label{fig: Figure 1}}
\end{figure}
\par\end{center}

\paragraph*{Results for RSET in Kerr.\textendash{} }

The numerical computation of the RSET is much more challenging. It
requires more modes and also higher accuracy for each mode, because
the divergence is stronger. As a consequence, our numerical results
for $\left\langle T_{\alpha\beta}\right\rangle _{ren}$ are less accurate
than for $\left\langle \phi^{2}\right\rangle _{ren}$. \footnote{The aforementioned oscillations in $\omega$ show up in the RSET computation
too, and we handle them (in both variants) just as we did in the calculation
of $\left\langle \phi^{2}\right\rangle _{ren}$.}

We point out that $\left\langle T_{\theta t}\right\rangle _{ren}$
and $\left\langle T_{\theta\varphi}\right\rangle _{ren}$ identically
vanish (mode by mode) for our massless  scalar field. In addition,
$\left\langle T_{rt}\right\rangle _{ren}$ and $\left\langle T_{r\varphi}\right\rangle _{ren}$
are individually-conserved components that do not require any regularization.
These two components are further addressed below. We shall refer to
the remaining six components, which do require regularization, as
``nontrivial''. Figure \ref{fig: Figure 2} displays results for
the six nontrivial components of $\left\langle T_{\alpha\beta}\right\rangle _{ren}$
in Unruh state, as functions of $r$, for two rays: $\theta=90^{\circ}$
and $\theta=0^{\circ}$. In the latter $\varphi$-splitting is invalid,
hence only $t$-splitting results are shown. At $\theta=90^{\circ}$
we provide results from both $t$-splitting and $\varphi$-splitting. 

We estimate the accuracy of the $t$-splitting results to be better
than one part in $10^{3}$ (throughout $r\ge2.5M$). The disagreement
between the two variants at the equator is at worst $\sim4\%$ (for
$r=2.5M$),  but it is usually better than $\sim1\%$, it improves
with increasing $r$, and at $r=10$ it is about one part in $10^{3}$.
Here again, the disagreement between the two variants is predominantly
associated to limited accuracy of $\varphi$-splitting. 

Figure \ref{fig: Figure 3} displays all six nontrivial RSET components
as functions of $\theta$ for $r=6M$. The $t$-splitting results
are estimated to be accurate to about one part in $10^{3}$. We also
present results from $\varphi$-splitting for $\theta\geq30^{\circ}$.
\textcolor{black}{The disagreement between the two is fairly large
(easily visible) at $\theta=30^{\circ}$, but it improves with increasing
$\theta$. It is a few percents for $\theta=35^{\circ}$ and reduces
to a few parts in $10^{3}$ at the equator.}

\begin{center}
\begin{figure}[h]
\centering{}\includegraphics[bb=25bp 180bp 540bp 650bp,clip,scale=0.45]{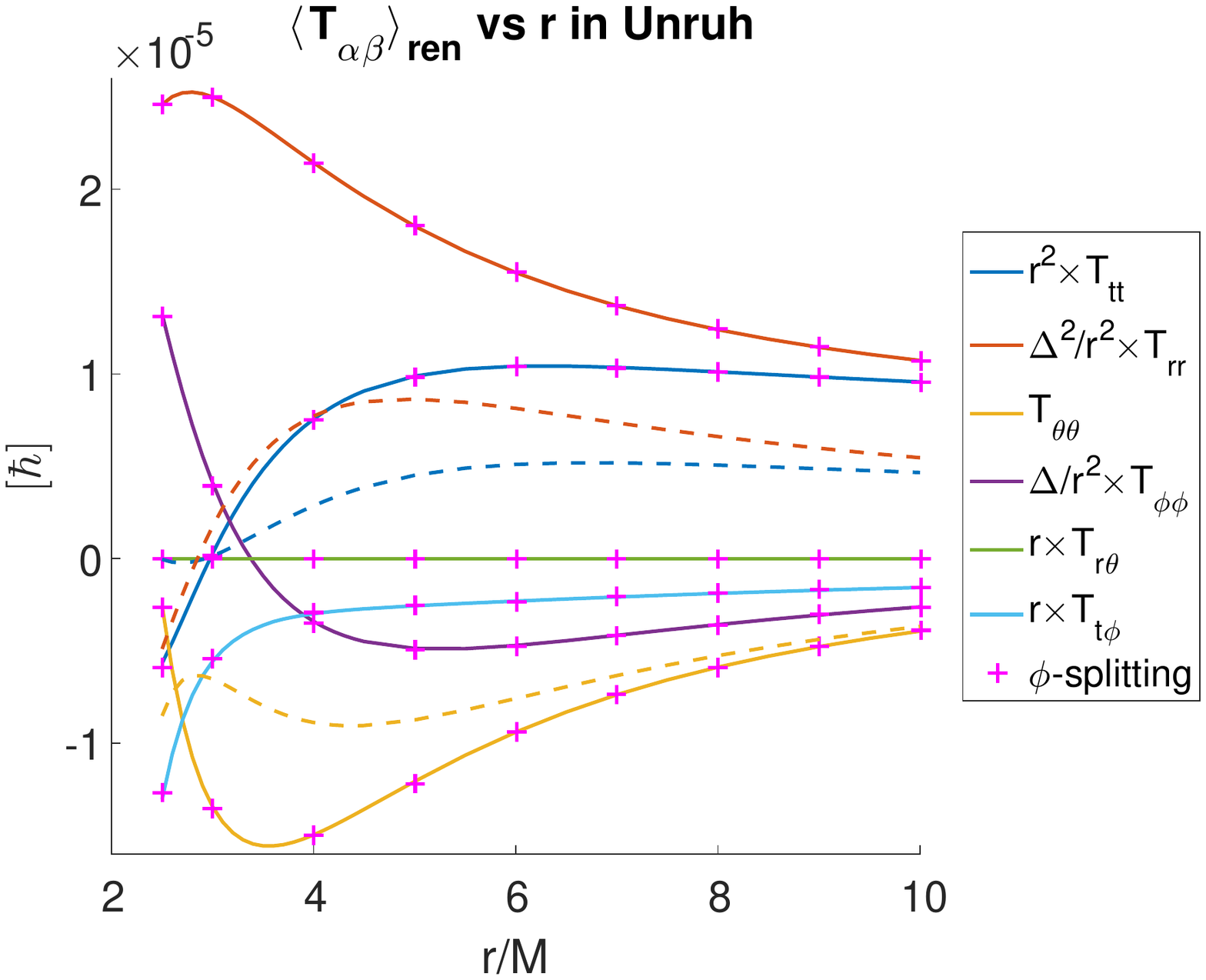}\caption{The six nontrivial RSET components as functions of $r$. The solid
curves and ``+'' symbols are results at $\theta=90^{\circ}$ from
$t$-splitting and $\varphi$-splitting respectively. The dashed curves
are $t$-splitting results at $\theta=0$. Notice that $\left\langle T_{r\theta}\right\rangle _{ren}$
(green line) vanishes at both the pole and equator, due to obvious
symmetry properties. Also, $\left\langle T_{\varphi\varphi}\right\rangle _{ren}=\left\langle T_{t\varphi}\right\rangle _{ren}=0$
at the pole.  \label{fig: Figure 2}}
\end{figure}
\par\end{center}

\begin{center}
\begin{figure}
\begin{centering}
\includegraphics[bb=25bp 180bp 540bp 650bp,clip,scale=0.45]{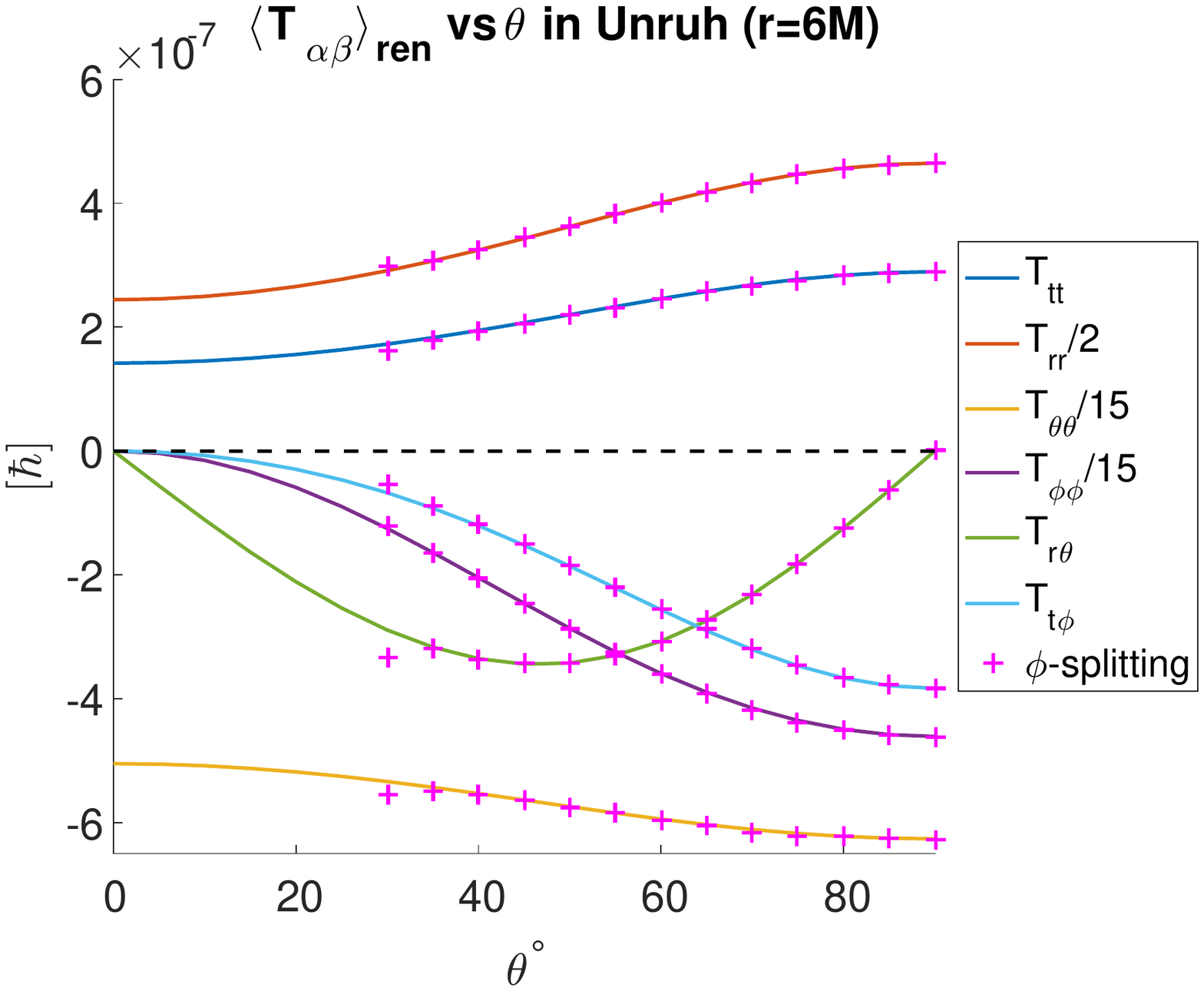}
\par\end{centering}
\caption{The six nontrivial RSET components at $r=6M$ as functions of $\theta$.
The solid curves and the ``+'' symbols are results obtained from
$t$-splitting and $\varphi$-splitting respectively. \textcolor{black}{The
deviations between the two are visible at $\theta=30^{\circ}$.} \label{fig: Figure 3}}
\end{figure}
\par\end{center}

\paragraph*{Energy-momentum conservation and conserved fluxes.\textendash{}}

An important consistency check of the computed RSET is energy-momentum
conservation $\left(\left\langle T^{\alpha\beta}\right\rangle _{ren}\right)_{;\beta}=0$.
In the PMR method one subtracts certain known tensors (derived from
Christensen's counter-terms) from the otherwise-divergent mode-sum.
We have analytically checked that these tensors are all conserved.
This ensures that the resultant RSET is conserved too, because the
contribution from the individual modes is guaranteed to be conserved.
We have also directly checked, numerically, the conservation of our
resultant RSET. \footnote{Due to the basic consistency of the different variants it is sufficient
to verify energy-momentum conservation in one of the variants. Here
we use $t$-splitting for this purpose, because it is technically
simpler (and numerically more accurate). }  

Two components of the conservation equation, $\alpha=t$ and $\alpha=\varphi$,
yield especially simple conservation laws:  

\[
\left\langle T_{rt}\right\rangle _{ren}=-K\left(\theta\right)/\Delta\,,\,\,\,\,\,\left\langle T_{r\varphi}\right\rangle _{ren}=L\left(\theta\right)/\Delta\,.
\]
The $r$-independent quantities $K\left(\theta\right)$ and $L\left(\theta\right)$
respectively represent the outgoing energy and angular-momentum flux
densities (multiplied by $r^{2}$), as measured by far observers placed
at various $\theta$ values. Note that these fluxes do not vanish
even in Boulware state, due to the Unruh-Starobinsky effect \cite{Starobinsky - Unrih starobinsky effect,Unruh - Unruh Starobinsky effect}.
Our results for $K\left(\theta\right)$ and $L\left(\theta\right)$
are displayed in Fig. \ref{fig: Figure 4}, for both Unruh and Boulware
states. Although the computation of $K\left(\theta\right)$ and $L\left(\theta\right)$
does not require regularization, to the best of our knowledge it is
the first time they are presented for a scalar field (results for
electromagnetic field are given in \cite{Ottewill - Kerr - electromagnetic differences}).

The integrals of $K\left(\theta\right)$ and $L\left(\theta\right)$
over the entire two-sphere yield the total energy flux $f$ and angular-momentum
flux $g$ emitted to infinity. In Unruh state we obtain $f=7.166\cdot10^{-5}\hbar/M^{2}$
and $g=1.8116\cdot10^{-4}\hbar/M$.  Of course these quantities can
also be directly computed using Hawking's original method \cite{Hawking - Particle creation by black holes},
which only requires numerical computation of the reflection and transmission
coefficients. This calculation (for a scalar field in Kerr) was done
by Taylor, Chambers and Hiscock \cite{Hiscock - Kerr evaporation}.
Our results agree with their computation to better than $1\%$, which
is their declared accuracy. To examine it more carefully we have repeated
their Hawking-radiation calculation, using MST method \cite{Maarten-MST}
for the reflection and transmission coefficients. We found agreement
better than one part in $10^{4}$ with the above mentioned results
for $f$ and $g$ (obtained from integrating $K,L$).

For Boulware state we obtain the integrated fluxes $f^{B}=1.265\cdot10^{-6}\hbar/M^{2}$
and $g^{B}=1.2187\cdot10^{-5}\hbar/M$, which express the Unruh-Starobinsky
effect.

\begin{center}
\begin{figure}
\begin{centering}
\includegraphics[bb=40bp 180bp 560bp 620bp,clip,scale=0.45]{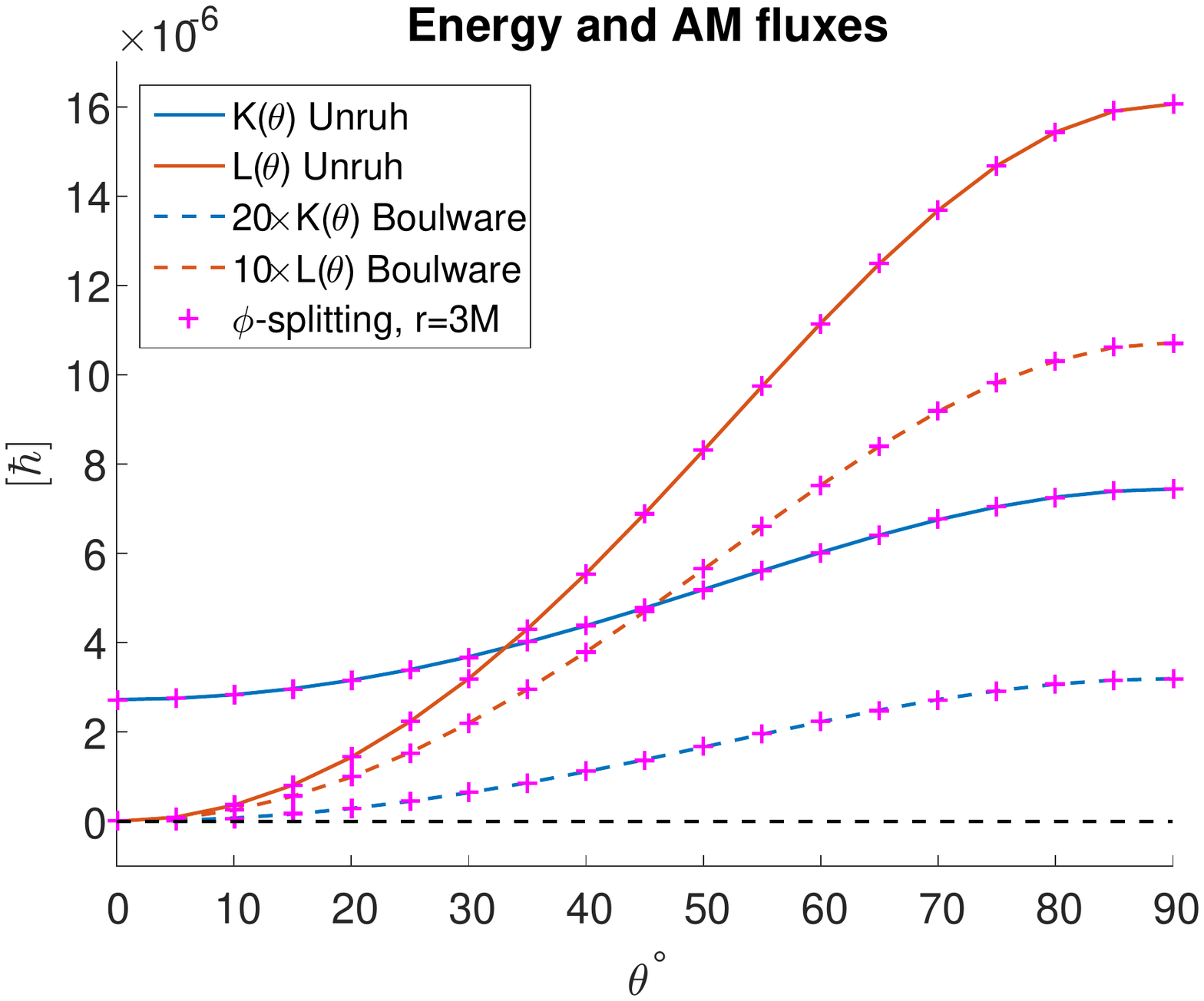}
\par\end{centering}
\caption{The  energy flux $K\left(\theta\right)$ and angular-momentum flux
$L\left(\theta\right)$. The solid curves are results in Unruh state,
and the dashed curves are in Boulware state \textemdash{} both obtained
using $t$-splitting at $r=10M$. The ``+'' symbols are results
obtained using $\varphi$-splitting at $r=3M$ (also confirming the
$r$-independence of $K$ and $L$). \label{fig: Figure 4}}
\end{figure}
\par\end{center}

\paragraph*{Discussion.\textendash{}}

We employed our PMR method to compute $\left\langle \phi^{2}\right\rangle _{ren}$
and $\left\langle T_{\alpha\beta}\right\rangle _{ren}$ in a spinning
BH, for a minimally-coupled massless scalar field, in both Unruh and
Boulware states. For brevity we mostly presented results for the more
realistic Unruh state, which represents physical evaporating BHs.
In addition, for Boulware state we displayed the fluxes of energy
and angular momentum to infinity (the Unruh-Starobinsky effect).

The regularization was done once using the $t$-splitting variant,
exploiting Kerr's stationarity, and once using the $\varphi$-splitting
variant, exploiting its axial symmetry, with good agreement between
the two variants in the regime where they both function properly. 

The usage of the two variants enables us to cross-check our results,
and it also helps assessing the numerical accuracy. In the domain
of $r$ and $\theta$ for which we have presented results, $t$-splitting
always provided more accurate results (This would change on approaching
the ergosphere boundary, where the temporal Killing field becomes
null). Note that the $\varphi$-splitting variant is more challenging,
as it requires two regularizations: the integral over $\omega$ (as
in $t$-splitting), and also the sum over $m$. One would of course
expect $\varphi$-splitting to fail at the polar axis, as the azimuthal
Killing field vanishes. We were a bit surprised to see that $\varphi$-splitting
starts to severely deteriorate fairly far from the pole: say, for
$r=6M$, at $\theta\lesssim15^{\circ}$ for $\left\langle \phi^{2}\right\rangle _{ren}$and
at $\theta\lesssim35^{\circ}$ for $\left\langle T_{\alpha\beta}\right\rangle _{ren}$.
\textcolor{black}{The situation improves with increasing $\omega_{max}$.}

One of the ultimate goals of the $\varphi$-splitting variant is to
allow investigating back-reaction effects in \emph{time-dependent,
spinning, evaporating BHs}. We see that in its present form $\varphi$-splitting
is not yet capable of achieving this goal, primarily due to the wide
domain of inapplicability around the pole. We do hope to improve this
situation.

Besides extending the computation to other $a/M$ values, and besides
improving the $\varphi$-splitting variant, we see several obvious
extensions of this work. The first is to compute the RSET inside the
ergosphere and also inside the horizon. This is important for understanding
how semiclassical effects would modify the internal structure of spinning
BHs. The second is to repeat our computation for other fields, e.g.
the electromagnetic field. Another possible use of our PMR method
is the RSET computation in the background metric of rapidly-rotating
relativistic stars (e.g. a neutron star).

We also hope that the ability to compute the RSET (and its corresponding
back-reaction) in a spinning BH will open the door for computing semiclassical
corrections to various physical phenomena. A possible example is corrections
to quasi-normal modes of evaporating spinning BHs \cite{Piedra - quasi normal modes,Andrei Belokogne and Antoine Folacci - Kerr Newman stress};
Another potential use for the RSET is in the context of the Kerr/CFT
corresponds \cite{Strominger - Extremal Kerr/CFT,Strominger - Kerr/CFT,Agullo Sals Olmo and Parker - Hawking radiation and Kerr/CFT,Strominger - CFT superradiance}. 

Full detail of this analysis of $\left\langle \phi^{2}\right\rangle _{ren}$
and $\left\langle T_{\alpha\beta}\right\rangle _{ren}$ in a Kerr
BH will be presented elsewhere.
\begin{acknowledgments}
\textit{Acknowledgment.\textendash{}} We are grateful to Leor Barack
for many fruitful discussions, and to Marc Casals for his helpful
advice. This research was supported by the Asher Fund for Space Research
at the Technion. MvdM was supported by the European Research Council
under the European Union's Seventh Framework Programme (FP7/2007-2013)
ERC grant agreement no. 304978. The numerical results in this paper
were obtained using the IRIDIS High Performance Computing Facility
at the University of Southampton.
\end{acknowledgments}

\end{document}